\documentclass{article}
\usepackage{LaThuileFPSpro, epsfig}
\newcommand{\GeVc}{\ensuremath{ {\rm GeV\!/}c   } }
\newcommand{\MeVcSq}{\ensuremath{ {\rm MeV\!/}c^2 } }
\newcommand{\invpb}{\ensuremath{ \rm pb ^{-1} } }
\newcommand{\Dplus}{\ensuremath{ D^{+} }}
\newcommand{\Dsplus}{\ensuremath{ D_s^{+} }}
\newcommand{\Dstarplus}{\ensuremath{ D^{*+} }}
\newcommand{\Dzero}{\ensuremath{ D^{0} }}
\newcommand{\ubarn}{\ensuremath{~\mu{\rm b} }}
\newcommand{\invpbarn}{\ensuremath{~{\rm pb^{-1}} }}
\begin{document}
\title{ 
  CHARM PHYSICS AT CDF II
  }
\author{
	Ivan K. Furi\'{c} \\
	(for the CDF II collaboration)\\
  {\em M.I.T., Fermilab - CDF - MS\#318, Batavia, IL 60510-0500} \\
}
\maketitle

\baselineskip=11.6pt

\begin{abstract}
The CDF II detector has the capability of triggering on displaced tracks.
Because of this ability, CDF II has accrued large samples of charmed meson
decays to fully hadronic final states in $64~\invpb$ of $p\overline{p}$
collision data gathered at $\sqrt{s} = 1.96 ~\rm{TeV}$. Using initial Run II
data samples, the production cross sections for $J/\psi$, $\Dzero$, $\Dplus$,
$\Dstarplus$ and $\Dsplus$ mesons have been measured. Ratios of branching
ratios for Cabibbo suppressed final states and $CP$ asymmetries in $D^0$ meson
decays have been studied. A measurement of the mass difference $m(\Dsplus) -
m(\Dplus)$ has been done, and a limit for the branching fraction of the FCNC
$\Dzero \rightarrow \mu^+ \mu^-$ decays has been set.
\end{abstract}
\newpage
%\baselineskip=14pt
%
%%==========================================================================%%
\section{Introduction}
%%==========================================================================%%
%
The CDF II detector\cite{TDR} is a major upgrade of the original CDF detector
which last took data in 1996. In Run I of the Tevatron, CDF made important
contributions to $B$ physics, providing some of the best measurements of
masses, lifetimes, mixing and branching ratios.

For charm physics results, the most important part of the upgrade are the new
integrated tracking system and the new trigger system. The integrated tracking
system consists of three silicon systems (L00 \cite{L00}, SVXII \cite{SVXII},
ISL \cite{ISL}) and a low-material, large-radius drift chamber (COT)
\cite{COT}. The detector has a brand new three-level trigger system.  The new
features of the trigger include triggering on muons with lower transverse
momenta, and triggering on displaced tracks and vertices \cite{SVT}.

%
%%==========================================================================%%
\section{$\bf J/\psi$ Production Cross Section}
%%==========================================================================%%
The mechanisms of $J/\psi$ production in $p \overline{p}$ collisions are not
well understood. Production cross sections from the assumed two major sources,
$b \rightarrow J/\psi X$ and direct prompt decays, were measured to be higher
than the initial theoretical predictions \cite{JPSITHEO}. Recent theoretical
advances in the extraction of the non-perturbative fragmentation functions of
the $B$ mesons from LEP data in a way that is consistent with the NLO QCD
calculations of the $b$ hadroproduction cross-sections have improved agreement
between theoretical predictions and CDF Run I $b \rightarrow J/\psi X$ cross
section measurements to better than 50 \%.

\begin{figure}
\epsfig{file = 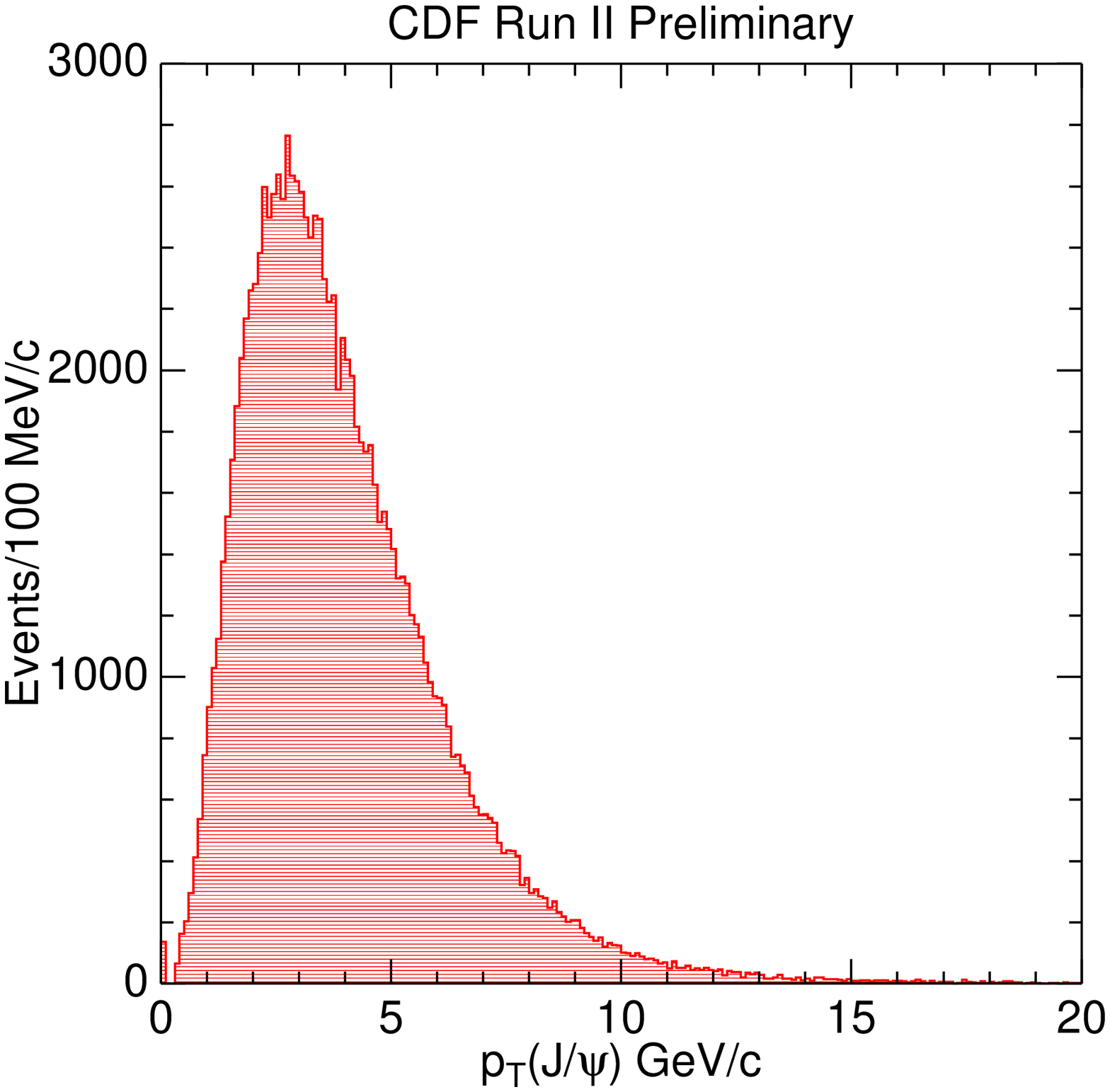,     width = 0.5 \textwidth} \nolinebreak
\epsfig{file = 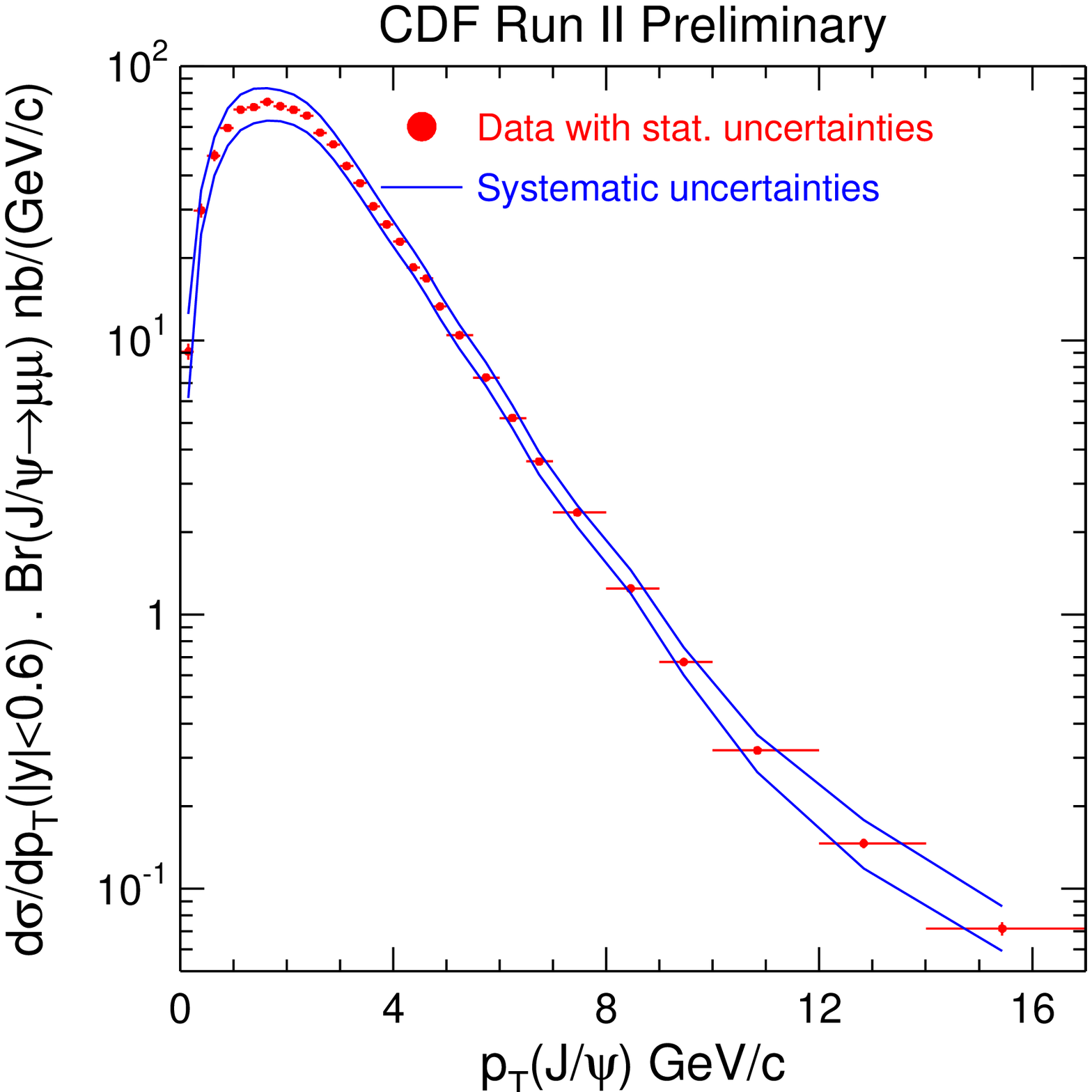, width = 0.5 \textwidth}
\caption{$J/\psi$ cross section measurement: the left plot shows the $p_T$ distribution
of the reconstructed $J/\psi$ mesons, and the right shows the differential production
cross section ($d\sigma / d p_T$).}
\label{fig:jpsixsec}
\end{figure}

The CDF II detector has an improved dimuon trigger with a lower $p_T$
threshold ($p_T > 1.4~\GeVc$). This has extended the low transverse momentum
range of triggered $J/\psi \rightarrow \mu^+ \mu^-$ down to $p_T(\mu^+ \mu^-)
\geq 0~\GeVc$. In $39.7~{\rm pb}$ of the initial Run-II data, 300~000 $J/\psi
\rightarrow \mu^+ \mu^-$ decays have been reconstructed. As shown in Figure
\ref{fig:jpsixsec}, the transverse momentum of the reconstructed $J/\psi$
extends to 0 \GeVc. After correcting for acceptance, trigger and
reconstruction efficiencies, an integrated production cross section of $240 \pm
1(stat)\,^{ + 35}_{ - 28} (syst)~{\rm nb}$ has been measured for $J/\psi$
mesons with $p_T(J/\psi) > 0~\GeVc$ and $|\eta(J/\psi)| < 0.6$.

%% \begin{figure}
%% \epsfig{file = ptjpsi_allphi.eps,     width = 0.5 \textwidth} \nolinebreak
%% \epsfig{file = xsec_syst1_bless3.eps, width = 0.5 \textwidth} \\
%% \epsfig{file = xsec_syst2_bless3.eps, width = 0.5 \textwidth} \nolinebreak
%% \epsfig{file = xsec_ptsqr_syst1_bless3.eps, width = 0.5 \textwidth}
%% \caption{$J/\psi$ cross section epsfigs}
%% \label{fig:jpsixsec}
%% \end{figure}

%%==========================================================================%%
\section{$\bf \Dsplus - \Dplus$ Mass Difference}
%%==========================================================================%%

\begin{figure}
\epsfig{file = 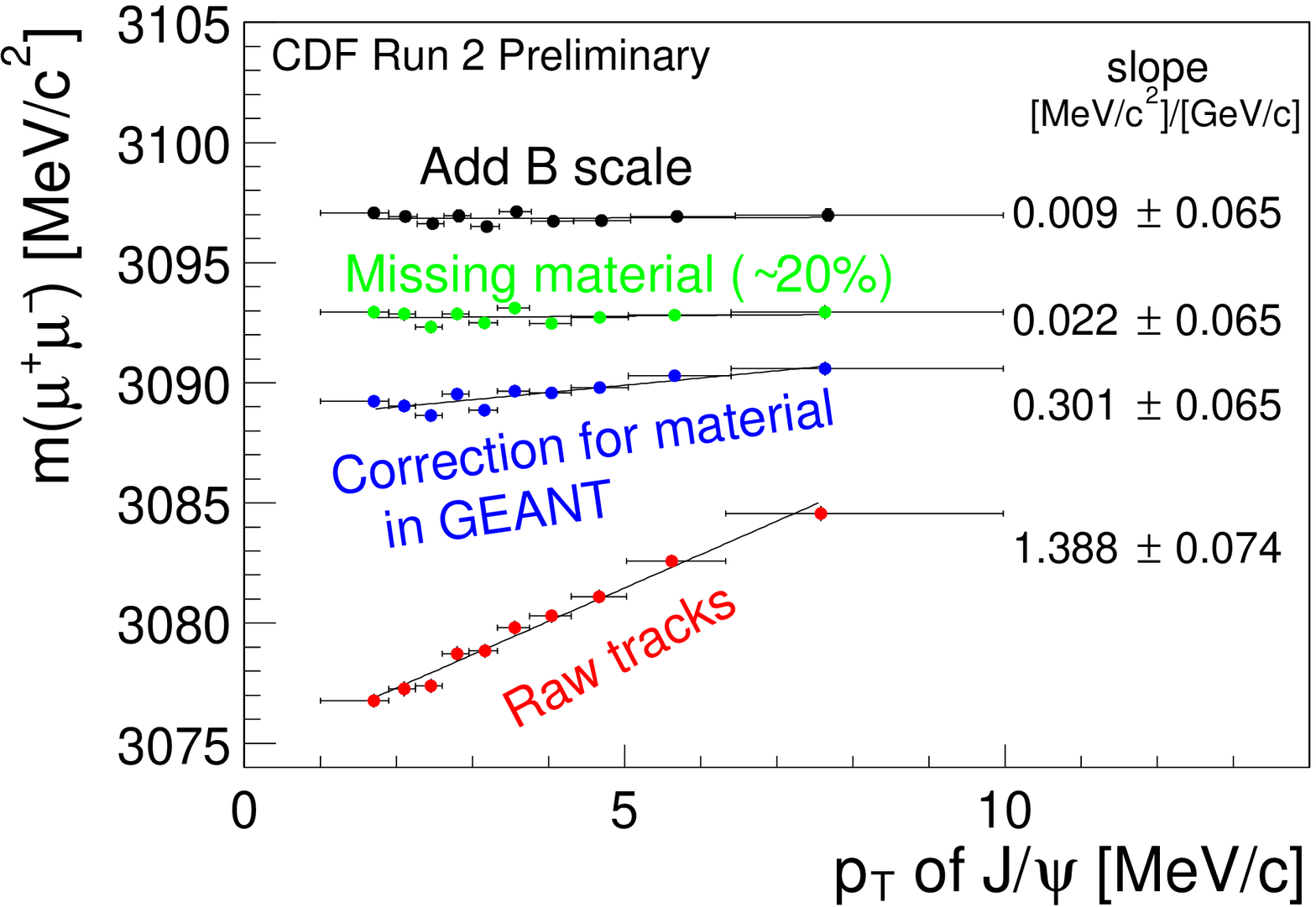,     width = 0.5 \textwidth, height = 0.47 \textwidth}
\epsfig{file = 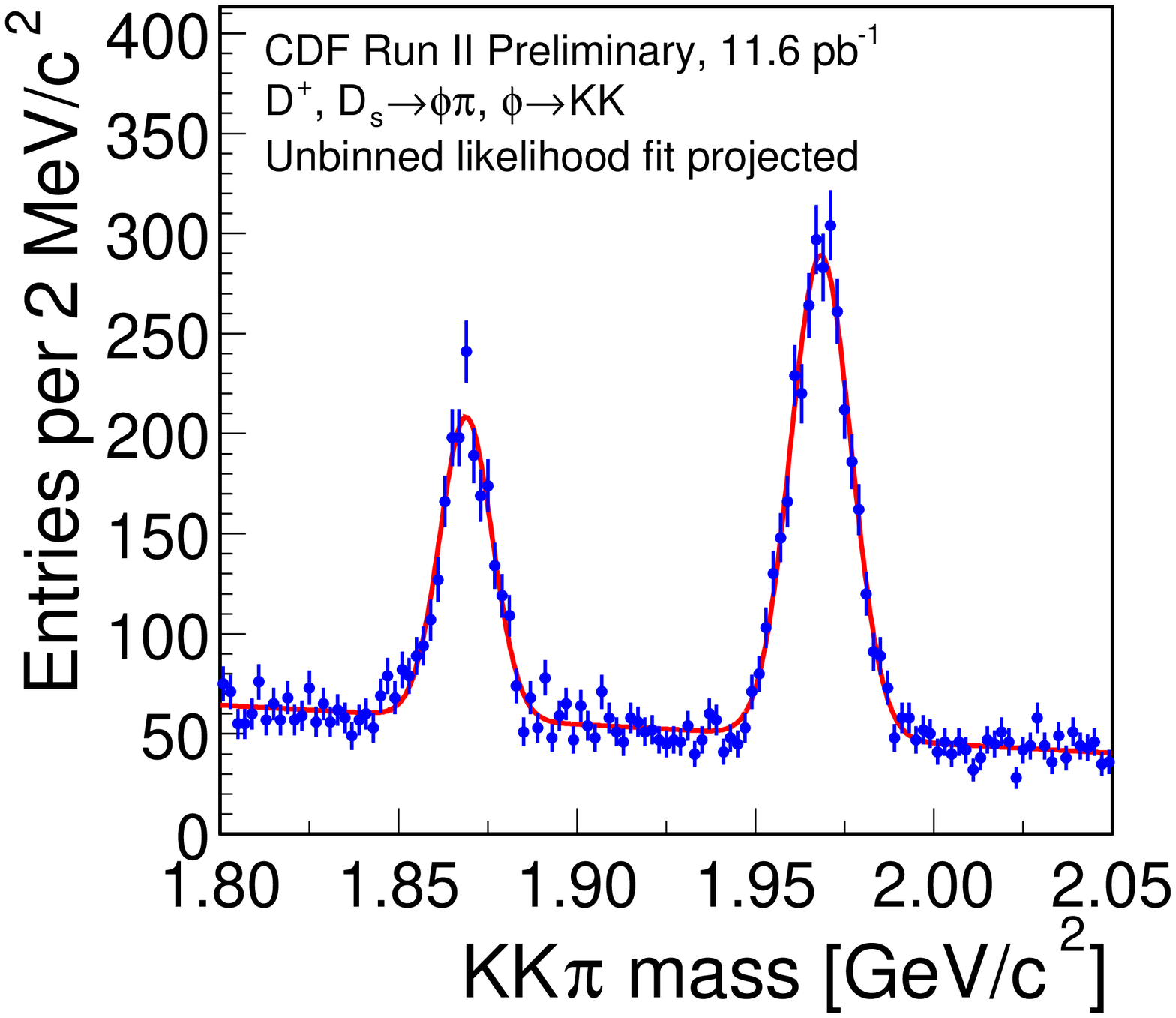, width = 0.5 \textwidth}
\caption{Momentum scale calibration and $m(\Dsplus) - m(\Dplus)$ mass measurement. The
left plot depicts the momentum scale calibration procedure, and the right
shows the invariant mass distribution for the $\Dsplus, \Dplus \rightarrow
\phi \pi$ signals with a superimposed fit.}
\label{fig:dsMassDiff}
\end{figure}

One of the first measurements done with the new sample of charmed meson decays
was the measurement of the mass difference $m(\Dsplus) - m(\Dplus)$. In a
sample corresponding to 11.6 $\invpb$, 2~400 $D_s \rightarrow \phi \pi$ and
$1~400$ $\Dplus$ decays were reconstructed. The detector invariant mass
resolution for these decays is about 8 \MeVcSq .  The momentum scale of the
detector was calibrated using 50~000 $J/\psi \rightarrow \mu^+ \mu^-$
decays. An outline of the procedure is depicted in Figure
\ref{fig:dsMassDiff}. The invariant mass of the $J/\psi$ decays shows a
dependance on the transverse momentum of the reconstructed $J/\psi$ because
the energy loss in the tracking system is not accounted for. After accounting
for energy loss according to the GEANT material map, a residual $p_T$
dependence can still be seen. Conversion scans of the tracking volume confirm
that there is material missing in the GEANT description so material is added
by hand to remove the $p_T$ dependence of the $J/\psi$ mass. The magnetic
field is scaled so that the $J/\psi$ mass agrees with the world average
\cite{PDG}. The results of the calibration (the amount of missing material and
the magnetic field) are cross-checked by reconstructing other charmed and
bottom meson decays ($\Dplus, \Dzero, \Upsilon$), and the reconstructed masses
are in good agreement with the corresponding world averages. This calibration
was then applied to the $\Dsplus, \Dplus \rightarrow
\phi \pi$ decays and the mass difference was found to be $m(\Dsplus) -
m(\Dplus) = 99.41 \pm 0.38 (stat) \pm 0.21(syst)~\MeVcSq$. The result is in
good agreement with previous measurements \cite{CLEODMass}, \cite
{BaBarDMass}. The systematic error is dominated by signal and background
modeling.

%%==========================================================================%%
\section{Charmed Meson Production Cross Sections}
\begin{figure}
\epsfig{file = 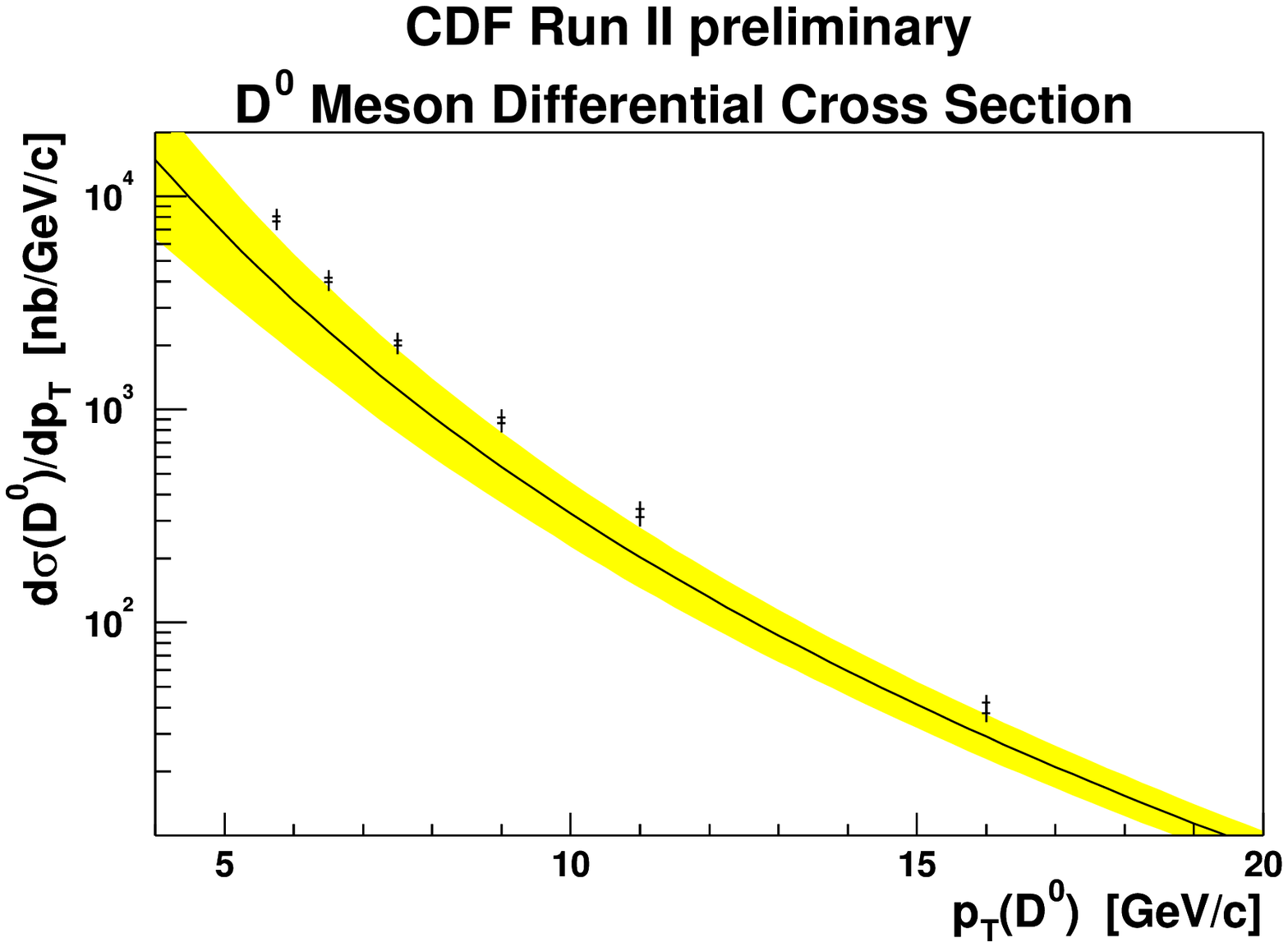,    width = 0.5 \textwidth, height = 0.5 \textwidth} \nolinebreak
\epsfig{file = 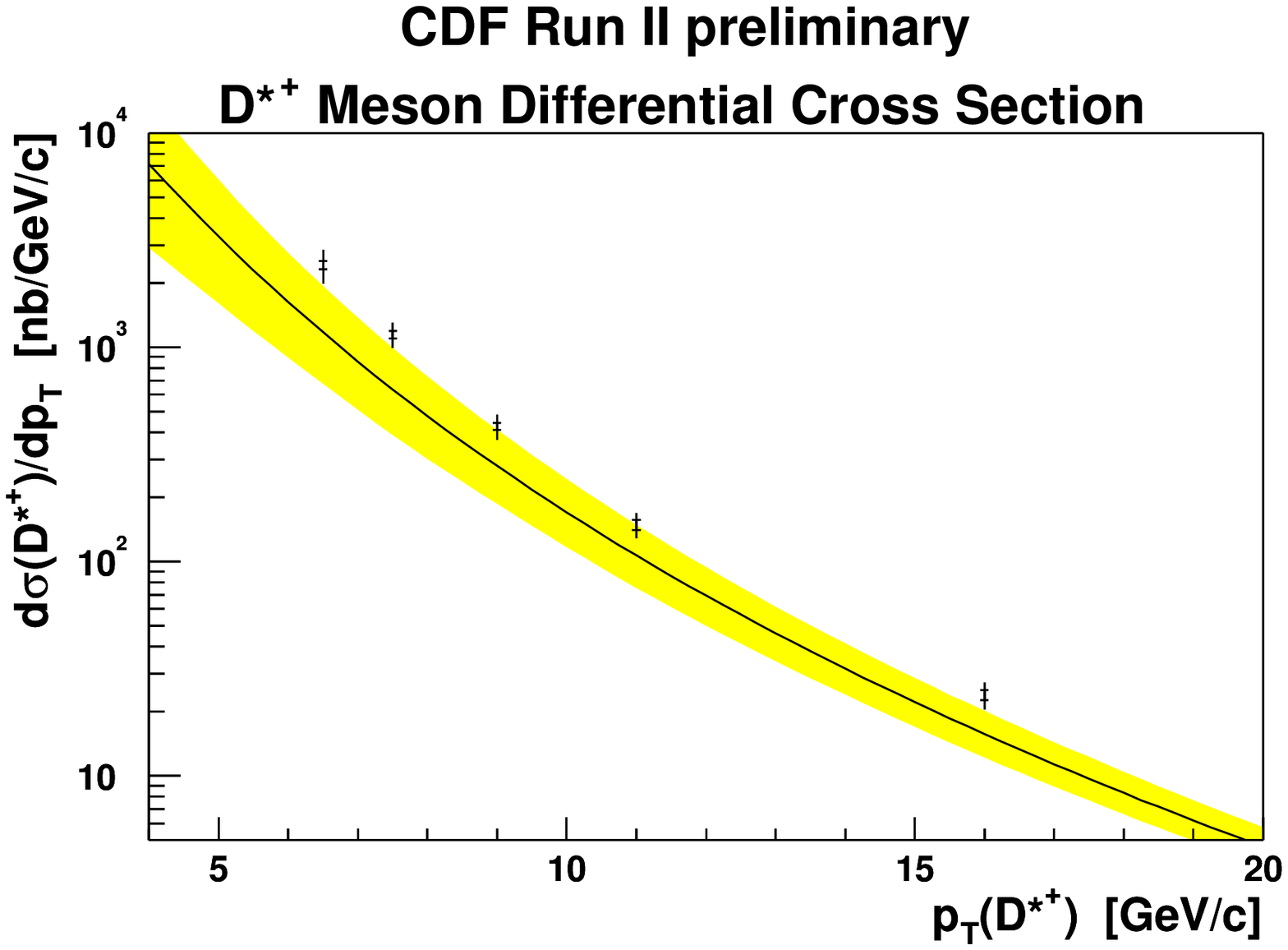, width = 0.5 \textwidth, height = 0.5 \textwidth} \\
\epsfig{file = 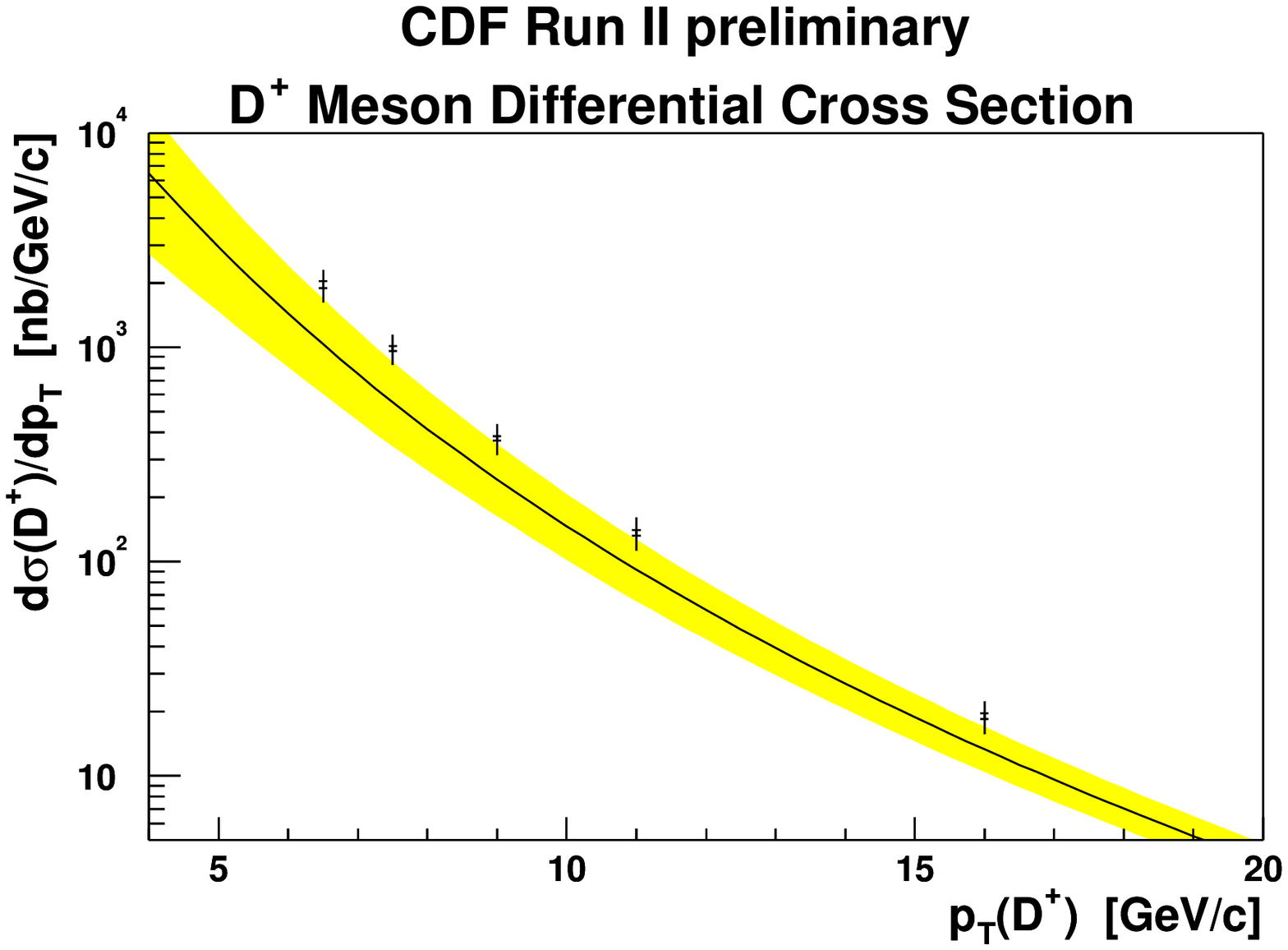,     width = 0.5 \textwidth, height = 0.5 \textwidth} \nolinebreak
\epsfig{file = 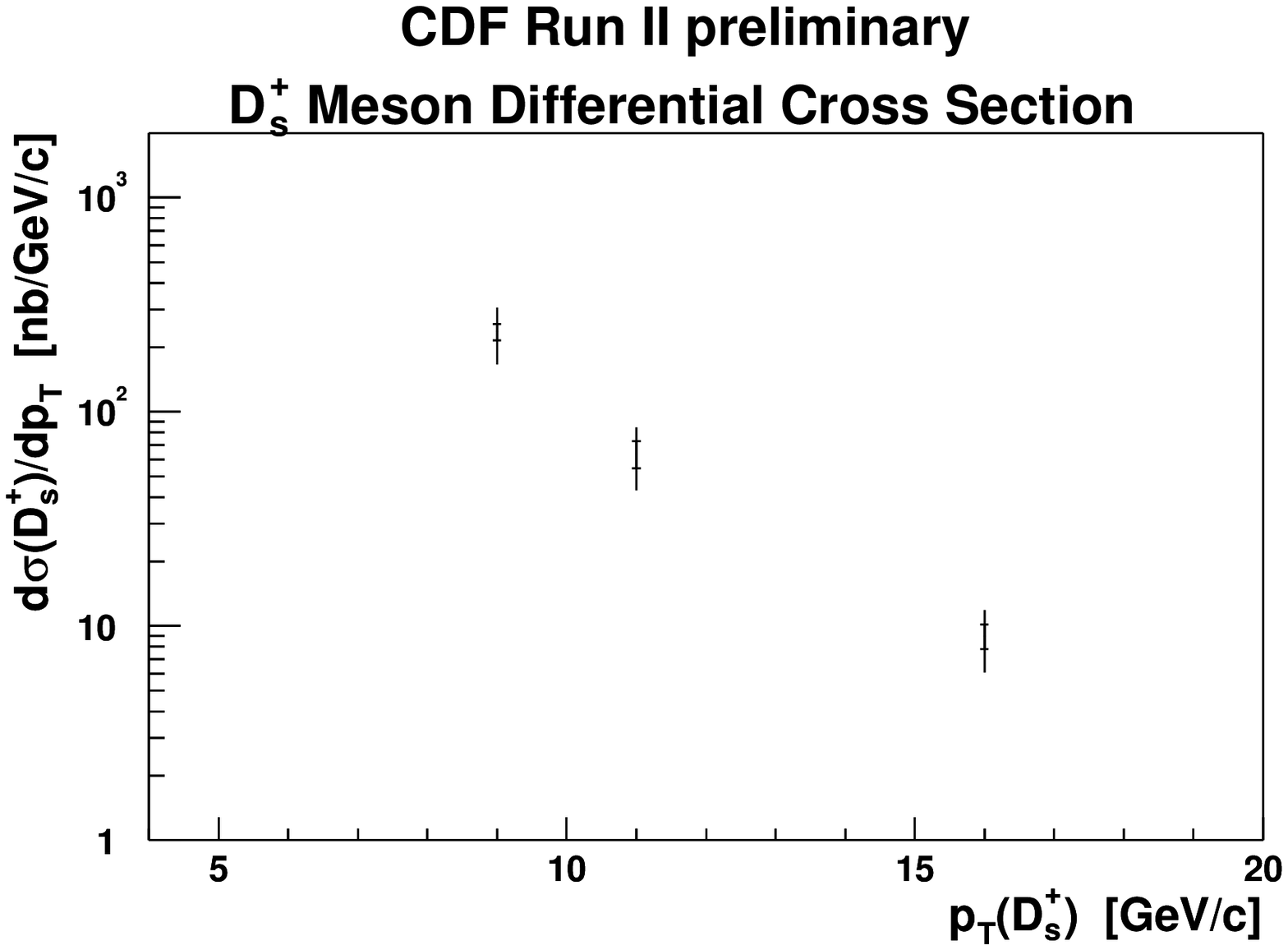,    width = 0.5 \textwidth, height = 0.5 \textwidth}
\caption{Charm meson differential cross sections for $\Dzero$, $\Dstarplus$, $\Dplus$
and $\Dsplus$ mesons, respectively. Theoretical predictions are overlaid upon the
measurement results.}
\label{fig:charmxsec}
\end{figure}

%%==========================================================================%%
In Run I, the B cross section for the $B^+ \rightarrow J/\psi K^+$ mode for
transverse momentum $p_T(B^+) > 6~\GeVc$ and rapidity $|y(B^+)| < 1$ was
measured to be $3.6 \pm 0.6 \ubarn$
\cite{CDFBXsec}. A preliminary measurement of the cross section for $\Dzero$, $\Dplus$, 
$\Dstarplus$ and $\Dsplus$ mesons was done with 5.7 $\invpb$ of Run II
data. The production cross sections were found to be larger than the
corresponding bottom meson cross sections: $4.3 \pm 0.1 (stat) \pm 0.7 (syst)
\ubarn$ for $\Dplus$, $9.3 \pm 0.1 (stat) \pm 1.1(syst) \ubarn$ for $\Dzero$
and $5.2 \pm 0.1 (stat) \pm 0.8 (syst) \ubarn$ for $\Dstarplus$ mesons. In the
case of the $\Dsplus$ mesons, the integrated production cross section was
measured for $p_T(\Dsplus) > 8~\GeVc, |y(\Dsplus)| < 1$ and found to be $0.75
\pm 0.05 (stat) \pm 0.22 (syst) \ubarn$. The differential production cross
sections ($d\sigma/d p_{T}$) for all the mesons are depicted in Figure
\ref{fig:charmxsec}.  Theoretical predictions (FONLL \cite{CXSECTHEO}) are
overlaid in the plots. The error band from the theory curve corresponds to the
maximum variation from changing the renormalization scale and the
factorization scales between $0.5$ and $2.0 \times \sqrt{p_T^2+m^2}$.

%%==========================================================================%%
\section{Branching Ratios and $CP$ Asymmetry}
\begin{figure}
\epsfig{file = 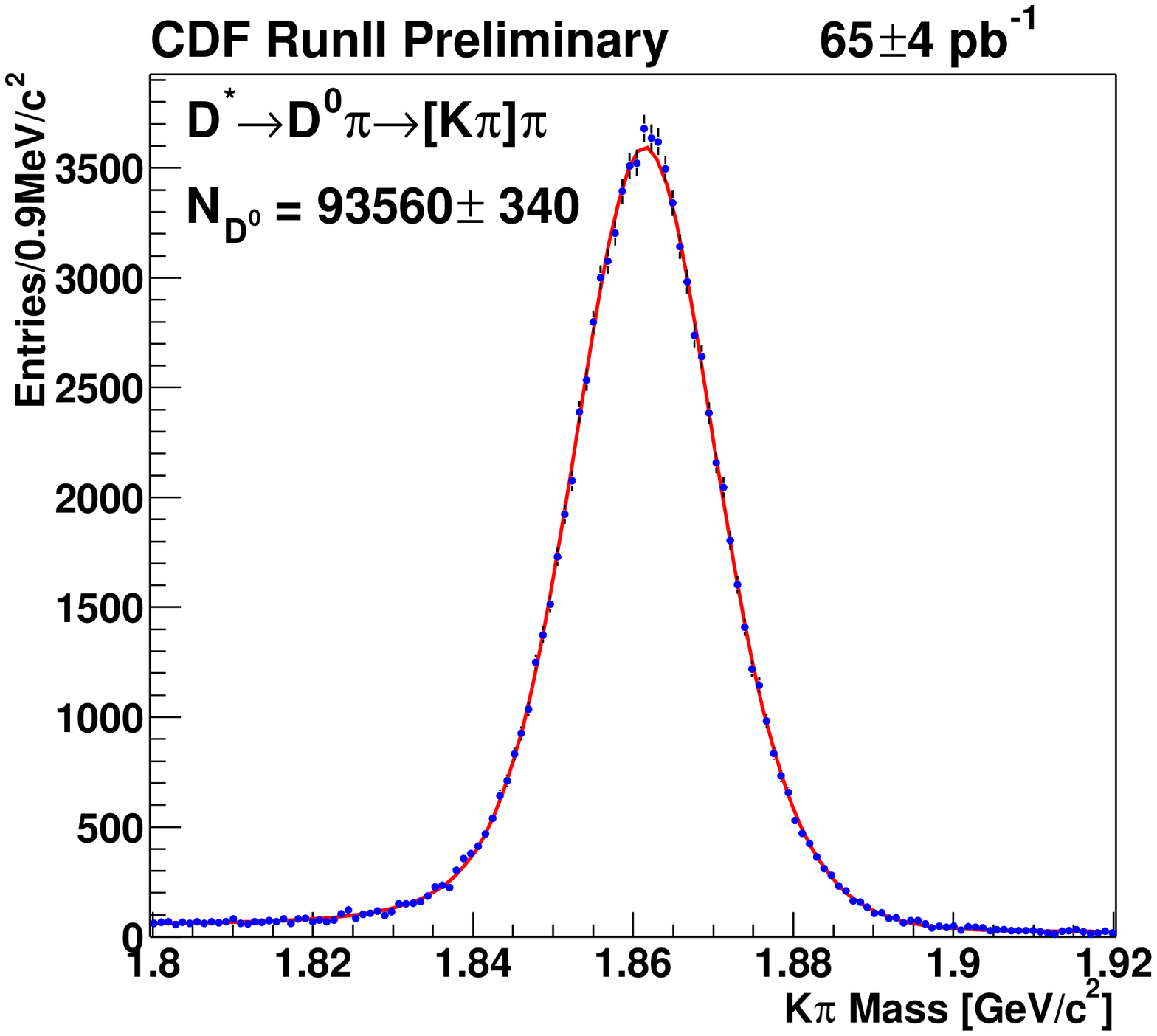,    width = 0.33 \textwidth}\nolinebreak
\epsfig{file = 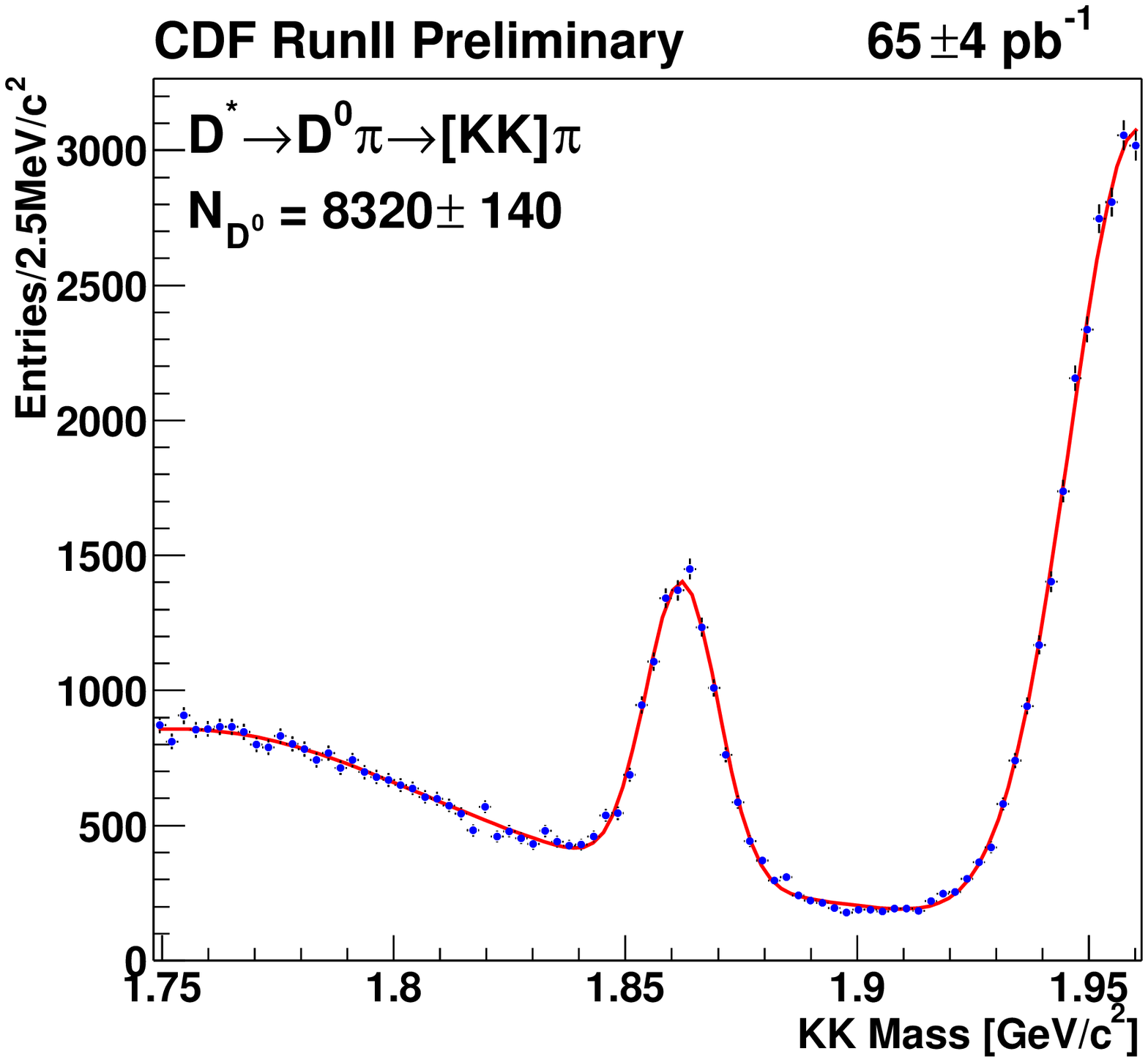,    width = 0.33 \textwidth}\nolinebreak
\epsfig{file = 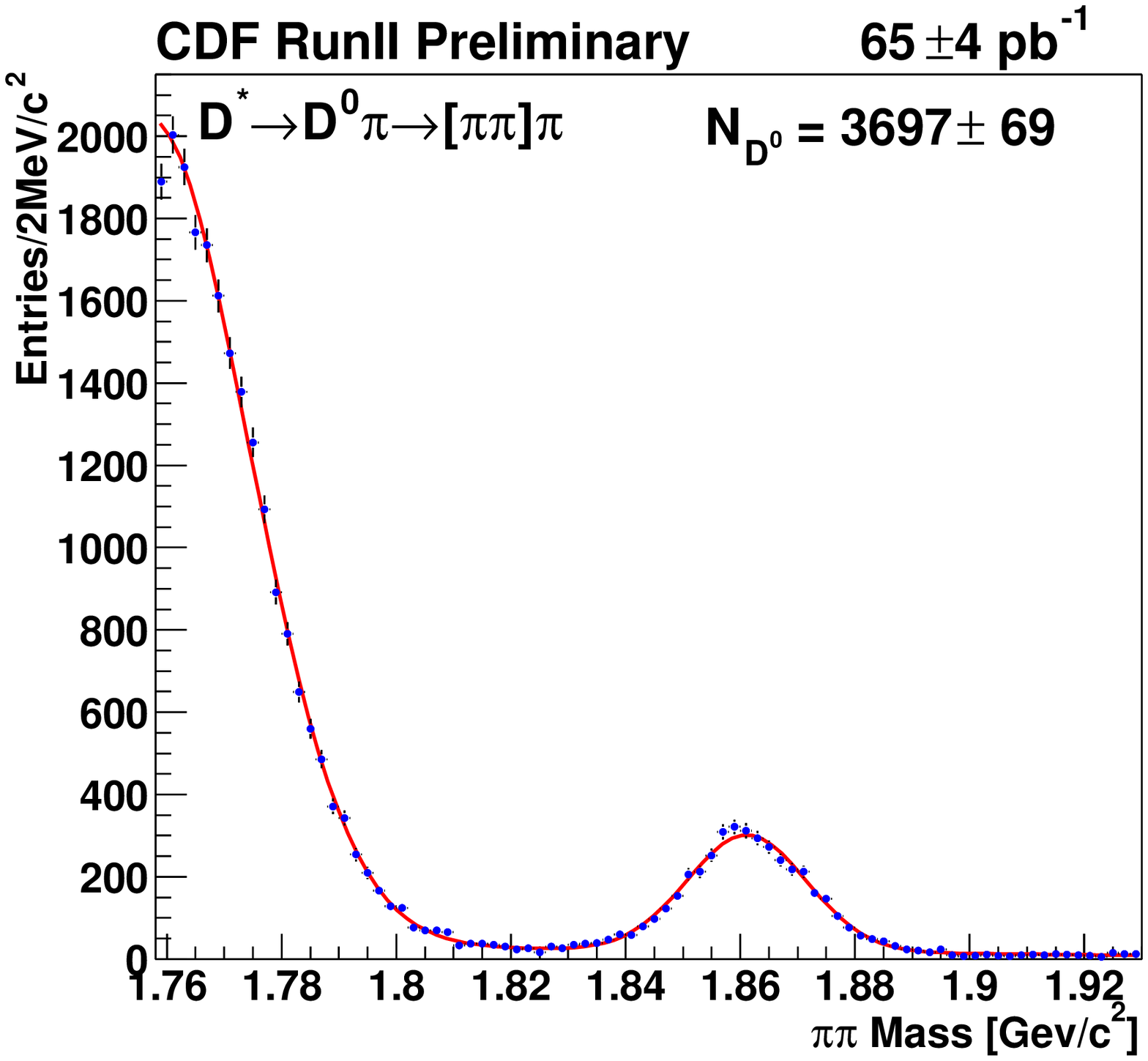,    width = 0.33 \textwidth}
\caption{Invariant mass distributions for the three two-body $\Dzero$ decays:
the dominant $\Dzero\rightarrow K^- \pi^+$, the Cabibbo suppressed $\Dzero \rightarrow K^+ K^-$
and $\Dzero \rightarrow \pi^+ \pi^-$. All three decay modes are reconstructed by requiring that 
the $D^0$ comes from the decay $\Dstarplus \rightarrow \Dzero \pi^+$.}
\label{fig:cabibbo-signals}
\end{figure}

%%==========================================================================%%
The study of the precise structure of the CKM matrix has been guided by
measurements of mixing and $CP$ violation in the neutral $K$ and $B$ meson
sectors.  The Standard Model predictions for the rate of mixing and $CP$
violation in the charm sector are small, with the predictions in both cases ranging
from 0.1\% to 1\% \cite{CPTHEO}. Observation of $CP$ violation above the 
$1\%$ level would be strong evidence for physics outside the Standard Model.
The SU(3) flavor symmetry predicts $\Gamma(D^0 \rightarrow K^+ K^-) /
\Gamma( D^0 \rightarrow \pi^+ \pi^-) = 1$ \cite{CSTHEO}, while the world average 
value is $2.88 \pm 0.15$ \cite{PDG}. This deviation is most likely caused by
large final state interactions (FSI)\cite{CSFSI}. In the initial $65 \pm 4
~\invpb$ of Run II data, 93~000 $\Dzero \rightarrow K^- \pi^+$, 8~300 $\Dzero
\rightarrow K^+ K^-$ and 3~700 $D^0 \rightarrow \pi^+ \pi^-$ decays were
reconstructed, as shown in Figure \ref{fig:cabibbo-signals}.  Good signal to
background was obtained by requiring that the $\Dzero$ always originates from
a $\Dstarplus$ decay: $\Dstarplus \rightarrow D^0 \pi^+$.  Using these samples
of $\Dzero$ decays, the following measurements of the ratios of branching
ratios were obtained by correcting the raw number of reconstructed candidates
by the relative trigger and reconstruction efficiencies::
\begin{equation}
\frac{ \Gamma ( D^0 \rightarrow K^+ K^- ) }{ \Gamma ( D^0 \rightarrow K \pi )} = 
9.38 \pm 0.18(stat) \pm 0.10 (syst)~\%
\end{equation}
\begin{equation}
\frac{ \Gamma ( D^0 \rightarrow \pi^+ \pi^- ) }{ \Gamma ( D^0 \rightarrow K \pi )} = 
3.686 \pm 0.076(stat) \pm 0.036 (syst)~\%
\end{equation}
The direct $CP$ asymmetries for $D^0$ decays were found to be $2.0 \pm 1.7 (stat) \pm 0.6 (syst) \% $
for $\Dzero \rightarrow K^+ K^-$ decays and $3.0 \pm 1.9 (stat) \pm 0.6 (syst) \% $ for $\Dzero \rightarrow \pi^+ \pi^-$ decays.

%% \begin{figure}
%% \epsfig{file = track_charge_asymmetry.eps,    width = 0.5 \textwidth}\nolinebreak
%% \epsfig{file = d0kpi_residual_asymmetry.eps,    width = 0.5 \textwidth}
%% \caption{$D^0$ impact parameter}
%% \label{fig:cabibbo-signals}
%% \end{figure}

%%==========================================================================%%
\section{Rare Charm Decays}
%%==========================================================================%%
For the flavor changing neutral current decay $\Dzero \rightarrow \mu^+
\mu^-$, the Standard Model predicts a branching ratio of $Br(\Dzero
\rightarrow \mu^+ \mu^-) \sim 10^{-13}$. The present experimental limit is
$Br(\Dzero \rightarrow \mu^+ \mu^-) \sim 4.1\times 10^{-6}$ from BEATRICE
\cite{BEATRICE} ($4.2 \times 10^{-6}$ from E771 \cite{E771}), 7 orders of
magnitude from the prediction.

New physics can substantially enhance this mode. In charm meson decays we are
constraining couplings to up-type quarks not necessarily constrained by $B$
decays. This makes $\Dzero \rightarrow \mu^+ \mu^-$ an unexplored region to
search for new physics.

\begin{figure}
\epsfig{file = 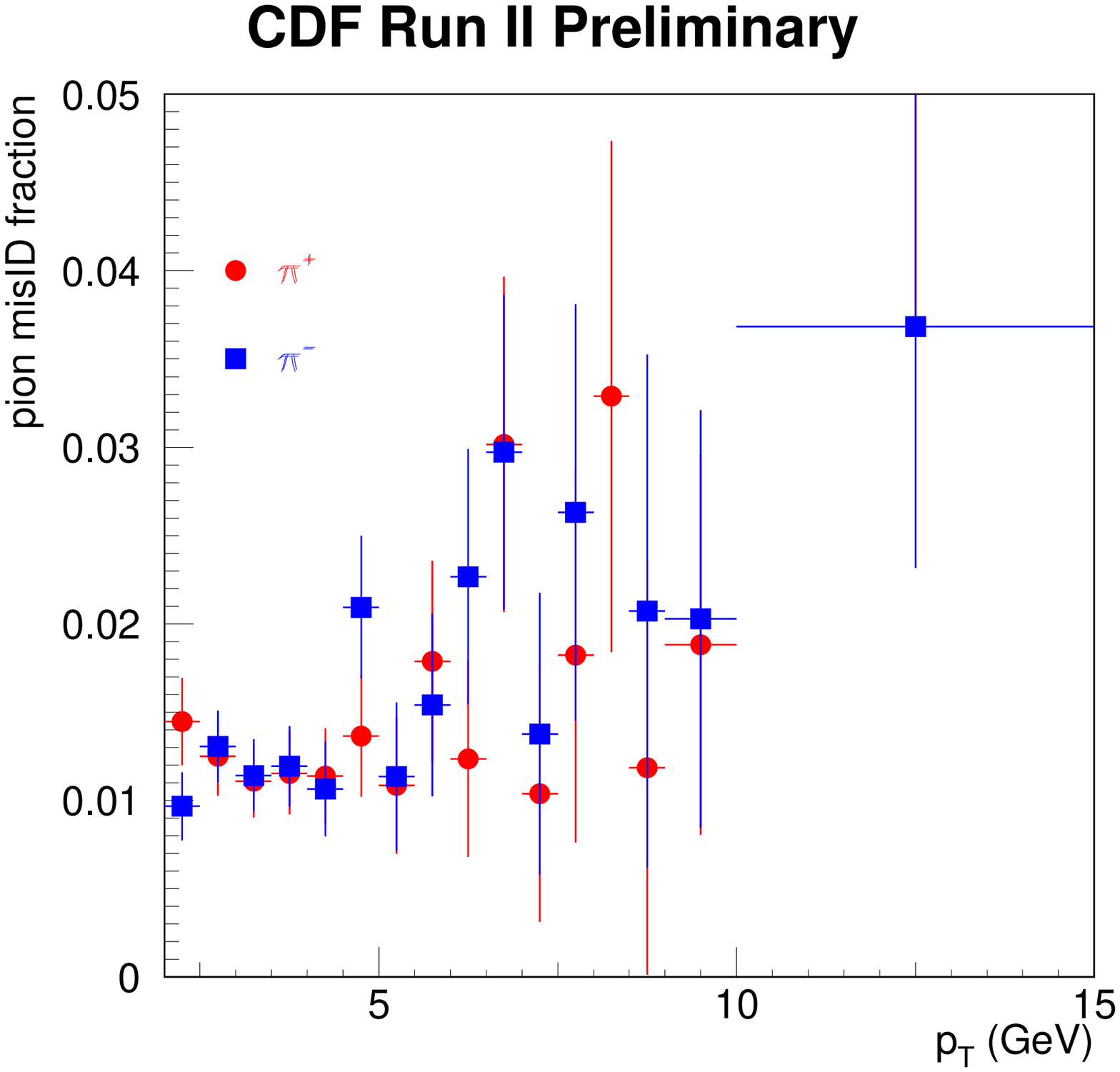 ,    width = 0.5 \textwidth}\nolinebreak
\epsfig{file = 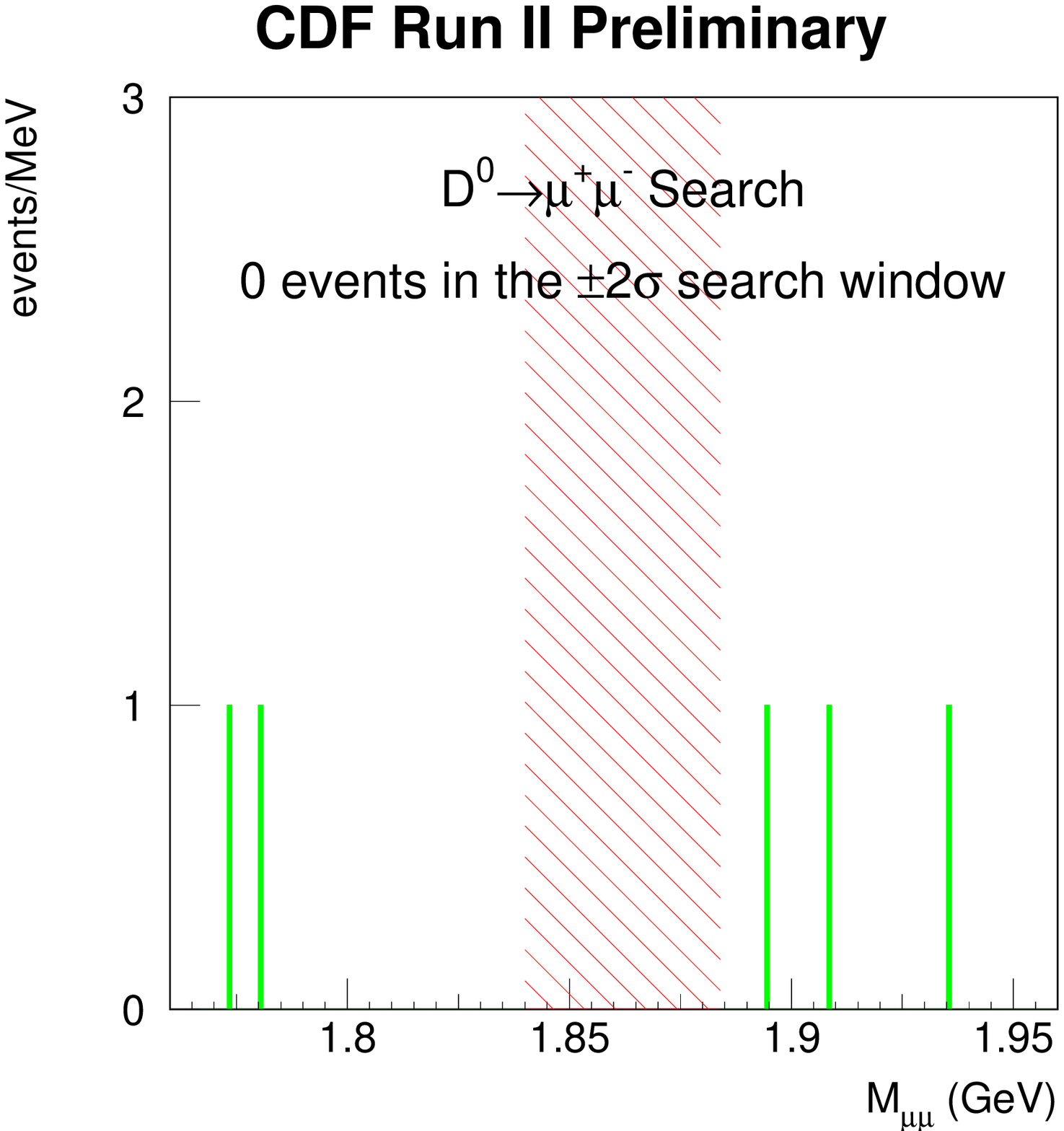,    width = 0.5 \textwidth}
\caption{Elements of the $\Dzero \rightarrow \mu^+ \mu^-$ analysis. The left plot
depicts the rate at which pions are misidentified as muons, and the right plot
shows the absence of events in the search window.}
\label{fig:rare-signals}
\end{figure}

Using $69 \invpbarn$ of Run II data, a search for $D\rightarrow \mu^+ \mu^-$
decays was performed.  As in the $CP$ asymmetry analysis, the $\Dzero$ decays
were reconstructed in a clean final state by requiring that they originate
from $\Dstarplus \rightarrow \Dzero \pi^+$ decays. The kinematically similar
$\Dzero \rightarrow \pi^+ \pi^-$ decay was used as a normalization mode. The
sources of background for this decay are $D^0 \rightarrow \pi^+ \pi^-$ events
in which both pions are misidentified as muons and combinatorial
background. Both the pion misidentification rate and the level of the
combinatorial background were measured from a kinematically similar but
statistically independant set of events. The estimated number of background
events in the search window was $1.7 \pm 0.7$. No events were found in the
search window, as seen in Figure \ref{fig:rare-signals},
and a limit was set on the branching ratio: $Br(\Dzero
\rightarrow \mu^+ \mu^-) < 2.4 \times 10^{-6}$ at 90\% C.L. This is currently
the worlds best limit on the $\Dzero \rightarrow \mu^+ \mu^-$ branching ratio.

%% \begin{figure}
%% \epsfig{file = mpipi_dstar.eps,    width = 0.5 \textwidth}\nolinebreak
%% \epsfig{file = mmumu_dstar.eps,    width = 0.5 \textwidth}
%% \caption{$D^0$ impact parameter}
%% \label{fig:rare-signals}
%% \end{figure}
%% 
%% 
%% \begin{figure}
%% \epsfig{file = pi_misid.eps,    width = 0.5 \textwidth}\nolinebreak
%% \epsfig{file = k_misid.eps ,    width = 0.5 \textwidth}
%% \caption{$D^0$ impact parameter}
%% \label{fig:rare-signals}
%% \end{figure}
%% 
%% 
%% \begin{figure}
%% \epsfig{file = sideband_dstar.eps,    width = 0.5 \textwidth}\nolinebreak
%% \epsfig{file = mumusearch.eps,    width = 0.5 \textwidth}
%% \caption{$D^0$ impact parameter}
%% \label{fig:rare-signals}
%% \end{figure}

%%==========================================================================%%
\section{Summary}
%%==========================================================================%%
The upgraded CDF II detector is back in operation and has gathered around 65
$\invpbarn$ of data which can be used for charm analyses. Due to its ability
to trigger on displaced tracks and vertices, the spectrum of charm results has
extended from $J/\psi \rightarrow \mu^+ \mu^-$ decays to include hadronic
decays of $\Dzero, \Dplus, \Dstarplus,$ and $\Dsplus$. With the modest amount
of data gathered so far, world class results have already been obtained.
%
%% %%==========================================================================%%
%% \section{Acknowledgments}
%% %%==========================================================================%%
%% %
%% Acknowledgements..
%% %

%
\end{document}